\documentclass[sigconf]{acmart}
\AtBeginDocument{%
  }

\setcopyright{cc}

\acmConference[MuC'26]{Mensch und Computer 2026 – Tagungsband, Gesellschaft für Informatik e.V.}{30. August - 02. September 2026}{Duisburg, Germany}
\acmDOI{10.18420/muc2026-mci-wip-337}

\usepackage{tabularx} 
\usepackage{todonotes} 
\usepackage{color} 
\usepackage{booktabs}

\usepackage{longtable}

\usepackage{placeins}
\presetkeys%
    {todonotes}%
    {inline, backgroundcolor=green!40}{}

\usepackage{pdfpages}

\begin{document}

\title{Designing and Evaluating Granular Consent for Data Sharing in Cardiac Disease Prevention}

\author{Pavithren V S Pakianathan}
\orcid{0000-0002-2232-3658}
\email{pavithren.pakianathan@lbg.ac.at}
\affiliation{%
  \institution{Ludwig Boltzmann Institute for Digital Health and Prevention}
  \city{Salzburg}
  \state{Salzburg}
  \country{Austria}
}
\affiliation{%
  \institution{LMU Munich}
  \city{Munich}
  \state{Bayern}
  \country{Germany}
}

\author{Rania Islambouli}
\orcid{0000-0003-1689-8938}
\email{rania.islambouli@lbg.ac.at}
\affiliation{%
  \institution{Ludwig Boltzmann Institute for Digital Health and Prevention}
  \city{Salzburg}
  \country{Austria}
}
\author{Magenta Jade Shipsey}
\email{shipsey.magenta@gmail.com}
\affiliation{%
  \institution{Ludwig Boltzmann Institute for Digital Health and Prevention}
  \city{Salzburg}
  \country{Austria}
}
\author{Laura Maaß}
\orcid{0000-0001-7354-8120}
\email{laura.maass@uni-bremen.de}
\affiliation{%
  \institution{SOCIUM Research Center on Inequality and Social Policy, University of Bremen}
  \city{Bremen}
  \country{Germany}
}
\affiliation{%
  \institution{Leibniz ScienceCampus Digital Public Health, University of Bremen}
  \city{Bremen}
  \country{Germany}
}

\author{Jan Smeddinck}
\orcid{0000-0003-0562-8473}
\email{jan.smeddinck@lbg.ac.at}
\affiliation{%
  \institution{Ludwig Boltzmann Institute for Digital Health and Prevention}
  \city{Salzburg}
  \country{Austria}
}

\renewcommand{\shortauthors}{V S Pakianathan et al.}
\newcommand\rev[1]{\textcolor{red}{#1}}
\begin{abstract}
Dynamic consent can promise end users with greater control, but little is known about how older adults with chronic conditions navigate the tradeoff between control and burden in granular consent mechanisms in health data life-cycles. Using a two-stage design process we evaluated this tradeoff. An expert workshop (n=5) informed the design requirements for granular dynamic consent prototype. We evaluated single step vs multi-step granularity in dynamic consent using prototypes with cardiac patients (n=7) using a mixed-methods study. Quantitative measures showed no significant differences between low- and high-granularity consent screens in usability, workload, perceived information control or willingness to share data. However, qualitative findings revealed a control–burden paradox and trust-dependent engagement with granularity. Participants sought greater transparency and control over AI-mediated data processing. We contribute implications for designing granular consent in health data life-cycles.
\end{abstract}




\maketitle

\section{Introduction}

The growing use of multi-modal health data from wearables and electronic health records is enabling new forms of clinical decision-making and biomedical research. However, the reuse of such data across contexts, often involving multiple actors and AI-driven analysis, challenges traditional one-time consent models \cite{blandfordHCIHealthWellbeing2019, andreottaAIBigData2021}. Dynamic consent has been proposed as an alternative paradigm, allowing individuals to revisit and \textit{negotiate} \cite{mortierHumanDataInteractionHuman2014} data-sharing preferences over time \cite{janekayeDynamicConsentPatient2015}. A central design consideration within this paradigm is \textit{granular consent}: the specificity with which users can express preferences across data types, recipients and purposes \cite{leeDynamicConsentSensorDriven2021}. In this paper we focus on granularity, which determines how expressive and how effortful  a consent interaction becomes. Yet granularity comes at a cost: as control becomes finer-grained, the interactional and cognitive burden placed on the user grows \cite{leeDynamicConsentSensorDriven2021}.

This tension is particularly relevant for older adults managing chronic conditions such as cardiovascular disease (CVD), who  routinely share continuous, complex and multi-modal data for primary and secondary use in healthcare settings, but may face higher sensitivity to cognitive effort, data literacy challenges, and uncertainty about data use \cite{krishnaswami_gerotechnology_2020, ijaz_role_2024, shao_determinants_2025}. Despite this, there remain open questions on how granular consent should be designed for chronic disease populations who encounter consent not as a single research decision but as a recurring part of their care pathway.


To address this gap, we conduct a two-stage user-centered study (Figure \ref{fig:study}). In Part 1, a co-design workshop with five interdisciplinary experts in digital health, ethics, and medical informatics elicited five design requirements for granular consent in CVD self-tracking. In Part 2, we operationalized these requirements into two prototype modalities, a single-step (low-granularity) and a multi-step (high-granularity) consent flow, and evaluated them with seven CVD patients aged 55–74 in a within-subject mixed-methods study combining standardized usability and workload measures with think-aloud and semi-structured interviews. 

We make three contributions:
(1) \textbf{Empirical}: We show that older CVD patients do not engage uniformly with granular consent, but selectively adjust their level of interaction based on trust in the recipient, perceived effort, and perceived risk. Additionally participants called for saliency of controls for AI processing in consent flows.
(2) \textbf{Design}: We identify design strategies for supporting flexible engagement with consent, balancing low-effort decisions with optional deeper control.
(3) \textbf{Theoretical}: We propose the \textit{Control–Burden Paradox} and \textit{context-sensitive (selectively enacted) granularity} as lenses for designing and reasoning about consent in chronic disease care.
\section{Related Work}

\subsection{The Health Data Lifecycle}

The expansion of mobile sensing and electronic health records has enabled large-scale collection and reuse of health data across clinical and research contexts \cite{kumarMobileWearableSensing2021, blandfordHCIHealthWellbeing2019}. While individuals are generally willing to share health data for societal and clinical benefit \cite{bietzOpportunitiesChallengesUse2016, silberPreregisteredVignetteExperiment2023,helouFactorsRelatedPersonal2021, seltzerPatientsWillingnessShare2019, bainesPatientPublicWillingness2024, trinidadPublicComfortSharing2020a} \cite{westPersonalizedVascularHealthcare2022}, this willingness is challenged by increasing data reuse, particularly through AI systems that operate beyond the scope of initial consent \cite{andreottaAIBigData2021}. Traditional one-time consent models are poorly suited to these conditions, as they cannot accommodate evolving data flows, actors, and purposes \cite{lugerTermsAgreementRethinking2013, vayenaHealthResearchBig2018}. Mortier's human-data interaction framework \cite{mortierHumanDataInteractionHuman2014} explains while sharing data with online systems, end users should have \textbf{(1)} legibility of data processing \textbf{(2)} agency of data flows and \textbf{(3)} negotiability of data sharing preferences based on changing contexts. Aligning with this framework, there is growing interest in more flexible consent mechanisms (e.g., Dynamic Consent \cite{leeDynamicConsentSensorDriven2021}) that allow individuals to retain control over how their data is used over time.  

\subsection{Granular and Dynamic Consent}
Granular consent enables individuals to specify preferences across multiple dimensions of data sharing, including data type, recipient, and purpose. This aligns with Nissenbaum’s theory of \textit{Contextual Integrity}, which frames privacy as appropriate information flow governed by contextual norms \cite{helennissenbaumPrivacyContextualIntegrity2004}. However, recent work highlights that purpose of use is a critical dimension that must be explicitly communicated \cite{malkinContextualIntegrityExplained2023}. For instance, a patient could be sharing their data to a doctor for the purpose of getting cared for (Primary Use) or sharing it with digital health researchers for the purpose of improving research in the field concerning their disease (Secondary Use) \cite{commission_european_2025}. While granular controls are particularly valued in sensitive contexts \cite{naeemFactorsAssociatedWillingness2022}, their practical use remains uneven. This raises an interesting challenge: granular and dynamic consent potentially increases control, but also increases load and task complexity \cite{leeDynamicConsentSensorDriven2021, leeDynamicConsentPrivacyAware2022}. 

\subsection{Dynamic consent}

Dynamic consent systems operationalize granular consent by allowing users to manage preferences over time enabling ongoing, flexible, and transparent engagement \cite{janekayeDynamicConsentPatient2015, teare_reflections_2021}. However, empirical studies highlight significant usability challenges. High levels of granularity can lead to consent fatigue, low engagement, and reliance on default settings \cite{leeDynamicConsentSensorDriven2021, leeDynamicConsentPrivacyAware2022}. Attempts to mitigate this -- such as automation, bundling, or trigger-action rules—often reduce user effort but may also obscure decision-making \cite{daniela.epsteinFinegrainedSharingSensed2013}. Currently, existing work provides limited insight into how users actually engage with granular consent in practice—particularly in health contexts involving complex, multi-modal data sharing. 

\subsection{Research Gap}

Despite increasing interest in granular and dynamic consent, empirical evidence on how users, especially those with chronic conditions, navigate the trade-off between control and cognitive burden remains limited \cite{spencer_patient_2016, karway_my_2022}. This gap is especially relevant in clinical populations, such as individuals with CVD (typically older adults), where data collection is continuous, multi-modal (e.g., ECG, physical activity, blood pressure etc.), and involves multiple health system stakeholders. Initiatives such as Smart FOX in Austria \cite{donsa2024smart} and Kaye \emph{et. al} \cite{janekayeDynamicConsentPatient2015} highlight the growing need for patient-mediated data sharing infrastructures that enable secondary use of health data while maintaining trust, transparency, and control. These developments emphasize the importance of designing consent mechanisms which are appropriate for the contextual needs of specific population groups.
\section{Method - Part 1 Workshop}
A 90-minute workshop was conducted online via Zoom, facilitated by a moderator and a note-taker. It was centered on the Modular Open Research (MORE) Platform, a mobile health sensing system for cardiac prevention research developed at the authors' institutions \cite{pakianathan_multi-stakeholder_2023}. It aimed to elicit design requirements for dynamic and granular consent in cardiovascular disease (CVD) research, specifically focusing on consent parameters, contextual adaptation, and data transparency for primary and secondary uses of patient-generated health data. Participants were five interdisciplinary experts (n=5) spanning academic and applied research roles in digital health, ethics, data science, and medical informatics, who had between 2 and 27 years of professional experience. They were based in Austria, Germany, and Brazil. Participants received a gift voucher worth 50 EUR as a reimbursement. 
Participants engaged in a three-part co-design process: (1) eliciting consent parameters using scenario-based prompts, (2) exploring and prioritizing design ideas using 5 CVD patient personas and affinity clustering, and (3) evaluating a prototype informed by literature review and reflecting on implementation challenges, including biomedical AI considerations. A pilot session was conducted with HCI researchers (n=3) to refine the procedure.

\section{Findings - Part 1 Workshop}
The experts emphasized that for the 50+ CVD demographic, control must be balanced with legibility, ensuring that the interface remains accessible. Subsequently, the first two authors synthesized the workshop findings into a set of core design requirements based on feasibility of integrating into the MORE platform: 
\begin{enumerate}
\item \textbf{Temporal Transparency.} Make data retention visible and configurable (e.g., expiry dates).
\item \textbf{Reciprocity.} Provide feedback loops (e.g., study updates, personal insights)
\item \textbf{Structured Granularity.} Organize consent hierarchically (e.g., purpose-based grouping) to balance overview and detail.
\item \textbf{Context-Sensitive Control.} Adapt data-sharing preferences based on context (e.g., recipient, situation), including optional automation.
\item \textbf{Risk Visibility.} Surface less obvious consequences of data sharing (e.g., downstream use, indirect impacts).
\end{enumerate}
\subsection{Part 2: Probe-Based Interviews with CVD Patients}
Using the design requirements from Part 1 workshop, we designed a prototype consent interface with two levels of granularity. We evaluated two prototype consent modalities -- prototypes are available as supplements --  to examine how granularity impacts perceived information control and cognitive load among self-tracking CVD patients (N=7). Participants (ages 55–74; 2 female, 5 male) were recruited via social media and previous mailing lists. Inclusion criteria required a CVD diagnosis, above 50 years old, and more than 1 year of self-tracking experience.

\paragraph{Study Design and Procedure}
In a within-subject, counterbalanced study (AB/BA order), participants interacted with:
\begin{enumerate}
\item \textbf{Condition A (Single-step):} Reduced granularity; entities and purposes presented on one screen without data–recipient mapping.
\item \textbf{Condition B (Multi-step):} High granularity; multi-screen flow enabling specific data–recipient mapping.
\end{enumerate}

The study used a smartphone-based Figma prototype and a vignette involving data sharing for primary and secondary use in cardiac care and research settings. Additionally printed prototypes were used for thinking-aloud activity and for annotation of feedback -- see Appendix. Consent wording was adapted from the \href{https://www.medizininformatik-initiative.de/sites/default/files/2020-11/MII_WG-Consent_Patient-Consent-Form_v1.6d_engl-version.pdf} {Medical Informatics Initiative consent template}. Insurance-related use was excluded to align with the European Health Data Space (EHDS) framework. Participants engaged in "think-aloud" protocols, completed quantitative measures (NASA-TLX \cite{hart_development_1988}, SUS \cite{brooke_sus-quick_1996}, 7-point Likert-based Perceived Information Control Subscale (PIC) \cite{ayalon_evaluating_2019}, and single item 7-point Likert-based willingness-to-share data question, and participated in a semi-structured interview regarding AI and data-sharing contexts. Interviews were informed by privacy calculus theory \cite{laufer1977privacy}, which attempts to explain how individuals weigh benefits and costs associated with disclosing personal information and in our context, self-tracked information which varying sensitivity such as physical activity minutes, sleep duration and quality and GPS location. Audio data and transcripts were pseudonymized. Audio data were transcribed in German using a locally deployed speech-to-text system and manually reviewed for accuracy. Data were analyzed using a hybrid thematic analysis approach following Braun and Clarke \cite{braun_using_2006}. An initial codebook was developed deductively from the research questions and workshop themes and inductively refined during analysis. A single researcher coded the data, and then translated the transcripts into English. Two researchers then went through all the codes and resolved discrepancies through discussions, and collaboratively developed themes.

\subsection{Findings - Part 2 Interviews}
\subsubsection{Quantitative Analysis} 
We compared the modalities using a 2x2 Aligned Rank Transform (ART) ANOVA with prototype (Single-step vs. Multi-step) as a within-subjects factor and presentation order as a between-subjects factor. Across all measures - NASA-TLX, SUS, PIC, and willingness to share health data, no significant differences were found between the single-step and multi-step conditions. Both prototypes achieved high usability (SUS > 84), and minimal average workload across NASA-TLX subscales (M < 2.5). PIC was numerically higher in the single-step condition (M=5.62,SD=1.11) as compared to multi-step (M=4.86,SD=1.72), while willingness to share data was identical across both conditions (M=5.00,SD=1.83).

The lack of significant quantitative differences suggests that for the participant demographic, the choice between single and multi-step granularity was perceived to be neutral regarding standard usability metrics and load. However, during the interview, when asked about which option they preferred, most (n=7) preferred the multi-step approach. To understand the nuanced mental models of the patients that influenced these ratings, we turn to the thematic analysis of the think-aloud transcripts.

\subsubsection{Qualitative Analysis}
All participants were broadly willing to share their health data for improving care and supporting research outcomes and were indifferent to duration of data storage by an entity, with one noting that they would be willing to share ``as long as it serves or helps to improve other clinical pictures'' (P3). However, their interaction preferences differed based on trust in the data recipient, data type, level of anonymity, and perceived need for time and effort.

\paragraph{Control-Burden Paradox}
Participants valued granular control while also seeking low-effort interaction with the consent mechanism. Most participants preferred multi-step over the single step approach. The multi-step modality supported reflection, while participants also noted that completing consent was time-dependent (P1). Some described the multi-step interface as supporting reflection through “compact nuggets” (P3), while others noted that it ``\textit{requires time and attention}'' (P2). One participant described the multi-step approach as ``\textit{easier to revise and control},'' despite being slower, as it allowed them to ``\textit{specify on each screen exactly how long [they were] granting access}'' (P4). This suggests that participants did not interact with the two modalities uniformly, but regulated their level of engagement based on perceived effort and time constraints.

\paragraph{Contextual Granularity}
Our findings show trust varied by entity and data type. Participants expressed lower trust towards insurers and policy makers. One participant noted that health data is a “\textit{gold mine for insurance}”  and was concerned about politicians/policymakers taxing them for unhealthy habits which could be risk factors for heart disease (P7). Similarly, several participants expressed concerns about surveillance and reacted negatively to the term ``\textit{authorities.}'' For P1, the word ‘authorities’ sounded ``\textit{like the police}'' while P4 added that “\textit{government agencies simply shouldn’t have access to health data … whether I weigh 120 kilos or maybe just 80 kilos}.” In contrast, participants reported high trust in researchers and healthcare providers and were more willing to share de-anonymized data for clinical decision-making and research. P3 particularly mentioned that it would be fine to share ``\textit{collection of population-based, anonymized health trends}'' with policy makers and several others specifically mentioned that they would prefer granular control for less trusted entities (e.g. policy makers and insurers). Overall, participants preferred more granular control under low trust, while relying on simplified or “shortcut” interactions for high-trust entities such as researchers and healthcare providers.

\paragraph{Need for Saliency of Controls for AI data processing}
Participants perceived AI as pervasive but insufficiently transparent. P2 had a perception that AI is already being used as soon as data is shared with researchers and healthcare professionals. They emphasized that ``\textit{there’s no way around it}'' (P4) and were uncertain about how it functioned, how the data was used to train AI systems.  To increase acceptance in clinical context and digital health research, participants acknowledged that there should be human-oversight and visibility over health data flows to AI systems. One participant described feeling “powerless” in the face of AI (P5) and called for greater control and clearer disclosure of AI use at the beginning of the consent process, a view echoed by others (P6, P7). 
\section{Discussion}

Our findings show that users do not consistently engage with granular consent. Instead, they regulate their level of interaction based on perceived effort, trust in the data recipient, and understanding of data use. Instead, granularity introduces a control–burden trade-off that shapes how and when users engage with consent mechanisms. While our multi-step consent aimed to increase granular control, participants were indifferent to single step or multi-step in terms of load and expressed that the time they had and perceived effort for configuring the consent affected their preferences. Although they wanted more control and agency \cite{mortierHumanDataInteractionHuman2014}, they also wanted to have less effort, resulting in a paradox. Our findings uncover the privacy calculus of participants which relies on situational heuristics influenced by effort and context.

Trust in data recipients played a central role in shaping consent behavior. High-trust actors, such as healthcare providers and researchers, enabled permissive decisions, while low-trust actors, such as policymakers and insurers, required more granular control. Our findings extend Nissenbaum’s \textit{Contextual Integrity} \cite{helennissenbaumPrivacyContextualIntegrity2004} by showing that users do not consistently evaluate all contextual parameters (e.g., actor, purpose, data type). Instead, these dimensions are selectively engaged depending on trust and effort. This aligns with recent work emphasizing purpose as a key factor \cite{malkinContextualIntegrityExplained2023}, but further suggests that even when relevant, such parameters are not always actively considered. 

\subsection{Design implications}
Based on our findings we suggest that consent is dynamic and flexible based on user preferences, allowing for configurable granularity enabling users to decide between low-effort and detailed decisions. Furthermore, defaults could be automated based on their trust levels towards entities to reduce load - however, the degree of automation \cite{sheridan2005human} required could vary and requires future investigations. With developments in GenAI, natural language interactions could enable such a dialogic approach allowing users to configure their consent.  Finally, given participants’ uncertainty around AI, granular consent systems should provide early and clear disclosure of AI-related data use.
\section{Conclusion}
While granular consent in dynamic consent aims to empower patients, our study suggests two key tensions among older adult CVD patients: control burden paradox and contextual-granularity. We show that patients prefer better control but with lesser choices and that their preferences for granularity are contextual -- based on recipient trust. Furthermore, with AI integration into data systems, patients prefer its controls to be salient. Ultimately for meaningful consent situated in the health data life cycle \cite{blandfordHCIHealthWellbeing2019}, designers should offer controls allowing patients to set their data-sharing preferences flexibly.
\section{Limitations and Future Work}
This study has some limitations. First, the expert workshop included a small sample (n=5) with primarily European perspectives, which may limit the diversity of design inputs. Second, the patient study involved a small sample of CVD patients (n=7) who were already engaged in digital health and willing to share data, potentially biasing results toward more tech-savvy and motivated individuals. Third, the consent interfaces were based on a predefined set of data types and were presented as Figma prototypes. Real-world systems may involve more extensive and heterogeneous data streams, requiring higher cognitive burden of granular consent. Future work should validate these findings with larger and more diverse populations, and examine how consent preferences evolve in real-world, longitudinal settings. In particular, investigating how users engage with more complex and dynamic data configurations could further inform the design of adaptive consent systems.



\bibliographystyle{ACM-Reference-Format}
\bibliography{sample-base}
\FloatBarrier
\clearpage

\appendix
\onecolumn
\section{Appendix: Detailed Study Flow}
\label{appendix:studyFlow}
\begin{figure}[h]
    \centering
    \includegraphics[width=0.7\linewidth]{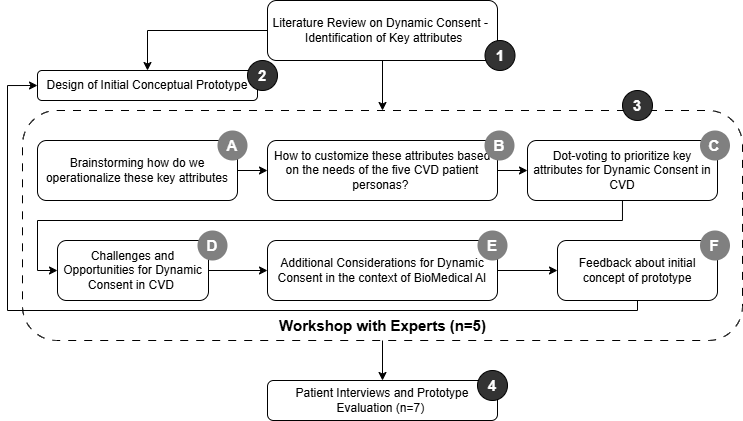}
    \Description{A flowchart detailing a four-stage process arranged vertically and connected by directional arrows. Stage 1 at the top center is labeled 'Literature Review on Dynamic Consent - Identification of Key attributes'. An arrow points left and down to Stage 2, labeled 'Design of Initial Conceptual Prototype'. Stage 1 and Stage 2 both point down into Stage 3, which is enclosed inside a large dashed rectangular box labeled at the bottom as 'Workshop with Experts (n=5)'. Inside Stage 3, six sub-steps marked A through F are arranged in two horizontal rows: The top row contains 'A. Brainstorming how do we operationalize these key attributes', pointing right to 'B. How to customize these attributes based on the needs of the five CVD patient personas?', pointing right to 'C. Dot-voting to prioritize key attributes for Dynamic Consent in CVD'. A line leads from C down and left to the bottom row, which contains 'D. Challenges and Opportunities for Dynamic Consent in CVD', pointing right to 'E. Additional Considerations for Dynamic Consent in the context of BioMedical AI', pointing right to 'F. Feedback about initial concept of prototype'. Arrow loops lead from F back to step D, as well as back up to Stage 2 outside the dashed box. Finally, an arrow leads from the bottom of the Stage 3 box down to Stage 4, labeled 'Patient Interviews and Prototype Evaluation (n=7)' at the bottom center.}
    \caption{Overall flow of the study}
    \label{fig:study}
\end{figure}

The research followed a four-stage process as illustrated in Figure \ref{fig:study}, involving literature synthesis, expert validation, and patient evaluation:

\begin{itemize}
    \item \textbf{Stage 1: Identification of Key Attributes.} The process initiated with a literature review on dynamic consent to identify the functional attributes necessary.
    \item \textbf{Stage 2: Initial Prototyping.} Findings from the literature were translated into an initial conceptual prototype.
    \item \textbf{Stage 3: Workshop with Experts ($n=5$).} A structured workshop was conducted to validate and refine the concept through several sub-activities:
    \begin{itemize}
        \item \textbf{(A) Operationalization:} Brainstorming how to translate abstract key attributes into functional system requirements.
        \item \textbf{(B) Persona Customization:} Tailoring attributes to the specific needs of five distinct CVD patient personas.
        \item \textbf{(C) Prioritization:} Using dot-voting to rank the importance of identified attributes.
        \item \textbf{(D) Challenges \& Opportunities:} Identifying systemic barriers to implementing dynamic consent in CVD care.
        \item \textbf{(E) BioMedical AI:} Specifically addressing considerations for data processing within AI-driven medical contexts.
        \item \textbf{(F) Feedback:} Collecting direct critiques of the initial prototype concept, which fed back into the design loop (Stage 2).
    \end{itemize}
    \item \textbf{Stage 4: Patient Interviews and Evaluation ($n=7$).} The final stage involved a qualitative evaluation of the refined prototype with patients to assess usability and trust.
\end{itemize}

\FloatBarrier
\newpage
\section{Appendix: Prototype Design}
\label{appendix:prototypeDesign}
\begin{figure}[h]
    \centering
    \includegraphics[width=\columnwidth]{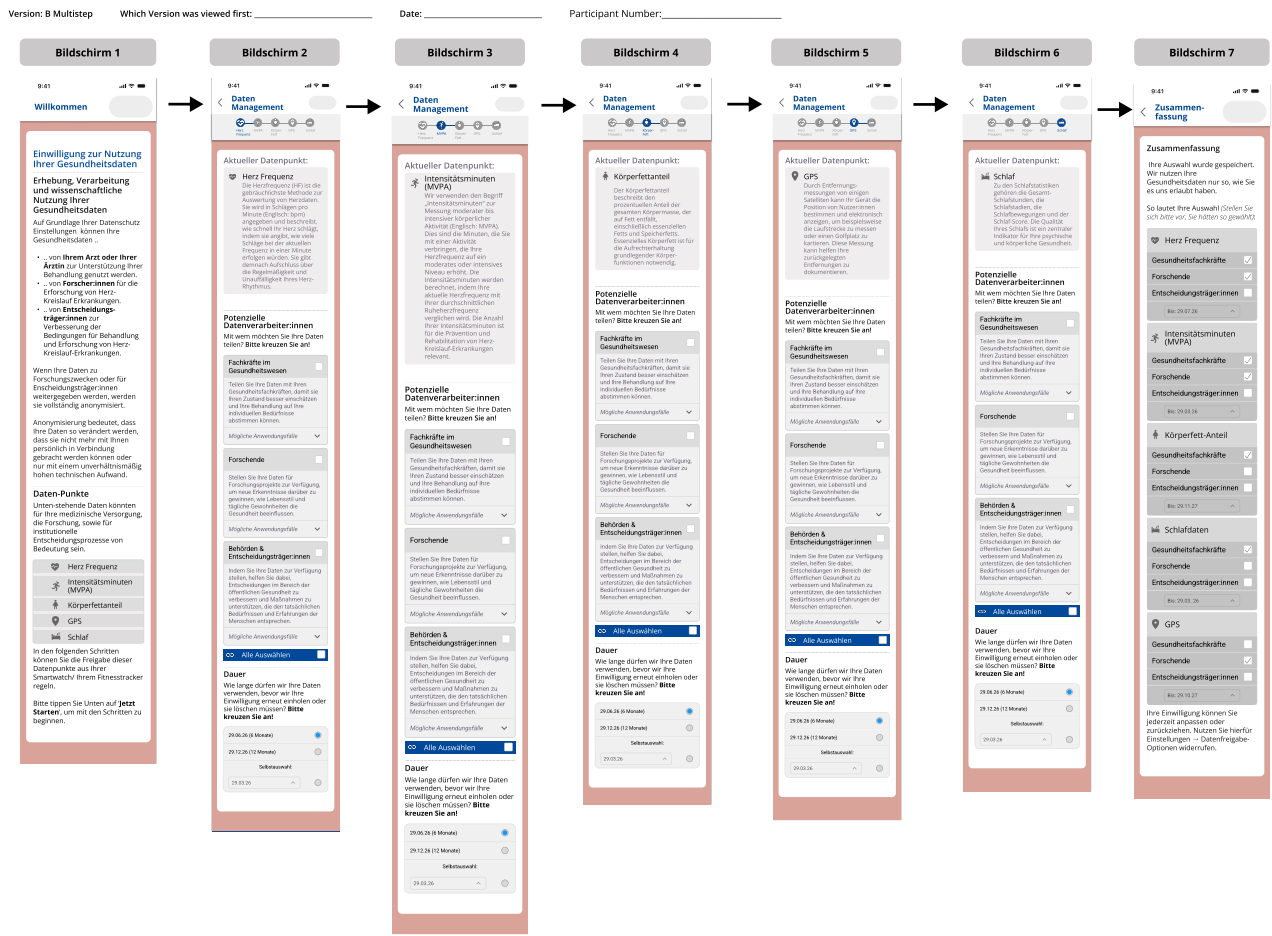}
    \Description{}
    \caption{Figma Prototype of Multi-Step Granular Dynamic Consent}
    \Description{A user interface flow titled showing a seven-screen mobile app sequence arranged horizontally from left to right, linked by right-pointing black arrows. Above the screens, metadata fields list 'Version: B Multistep', 'Which Version was viewed first:', 'Date:', and 'Participant Number:'. Each smartphone wireframe displays a light gray header bar and a main content region highlighted in translucent pink. Screen 1, labeled 'Bildschirm 1', presents a welcome screen titled 'Willkommen' and 'Einwilligung zur Nutzung Ihrer Gesundheitsdaten' with introductory text and bullet points outlining consent terms and dynamic data points. Screens 2 through 6, labeled 'Bildschirm 2' to 'Bildschirm 6', are titled 'Daten Management' at the top with a step-indicator bar containing circular nodes. Each of these five screens focuses on a specific health data metric listed under 'Aktueller Datenpunkt': Screen 2 describes 'Herz Frequenz' (Heart Rate), Screen 3 describes 'Intensitätsminuten (MVPA)' (Moderate-to-Vigorous Physical Activity), Screen 4 describes 'Körperfettanteil' (Body Fat Percentage), Screen 5 describes 'GPS' (Location Data), and Screen 6 describes 'Schlaf' (Sleep Data). Below each metric description, interactive cards allow users to select recipient categories under 'Potenzielle Datenverarbeiter:innen' (such as healthcare professionals, researchers, and public health decision-makers) via checkboxes, along with consent duration options listed under 'Dauer'. Screen 7, labeled 'Bildschirm 7', displays a final summary screen titled 'Zusammenfassung' listing all five metrics with checked user preferences and a concluding confirmation notice.}
    \label{fig:multi}
\end{figure}
\begin{figure}[h]
    \centering
    \includegraphics[width=0.7\linewidth]{images/SingleStepAnonymized.pdf}
    \caption{Figma Prototype of Single Step Dynamic Consent}
    \Description{A user interface flow showing a consolidated dynamic consent prototype consisting of two main mobile screens labeled 'Bildschirm 1' and 'Bildschirm 2', connected from left to right by a black arrow. On the left, 'Bildschirm 1' displays a long scrolling page under the header 'Daten Management'. The left section of this screen details consent conditions ('Einwilligung zur Nutzung Ihrer Gesundheitsdaten') followed by a list of health data metrics ('Daten-Punkte') including Heart Rate ('Herz Frequenz'), Activity Minutes ('Intensitätsminuten'), Body Fat ('Körperfettanteil'), Location ('GPS'), and Sleep ('Schlaf'). To its immediate right, an adjacent column within 'Bildschirm 1' details data recipient choices ('Potenzielle Datenverarbeiter:innen') such as healthcare professionals, researchers, and public decision-makers, along with consent duration options ('Dauer'). An arrow points right to 'Bildschirm 2', titled 'Zusammenfassung' (Summary), which displays a single mobile screen listing the selected health data categories, chosen recipient groups, expiration date options, and a final confirmation note.}
    \label{fig:single}
\end{figure}
\begin{figure}[h]
    \centering
    \includegraphics[width=0.7\linewidth]{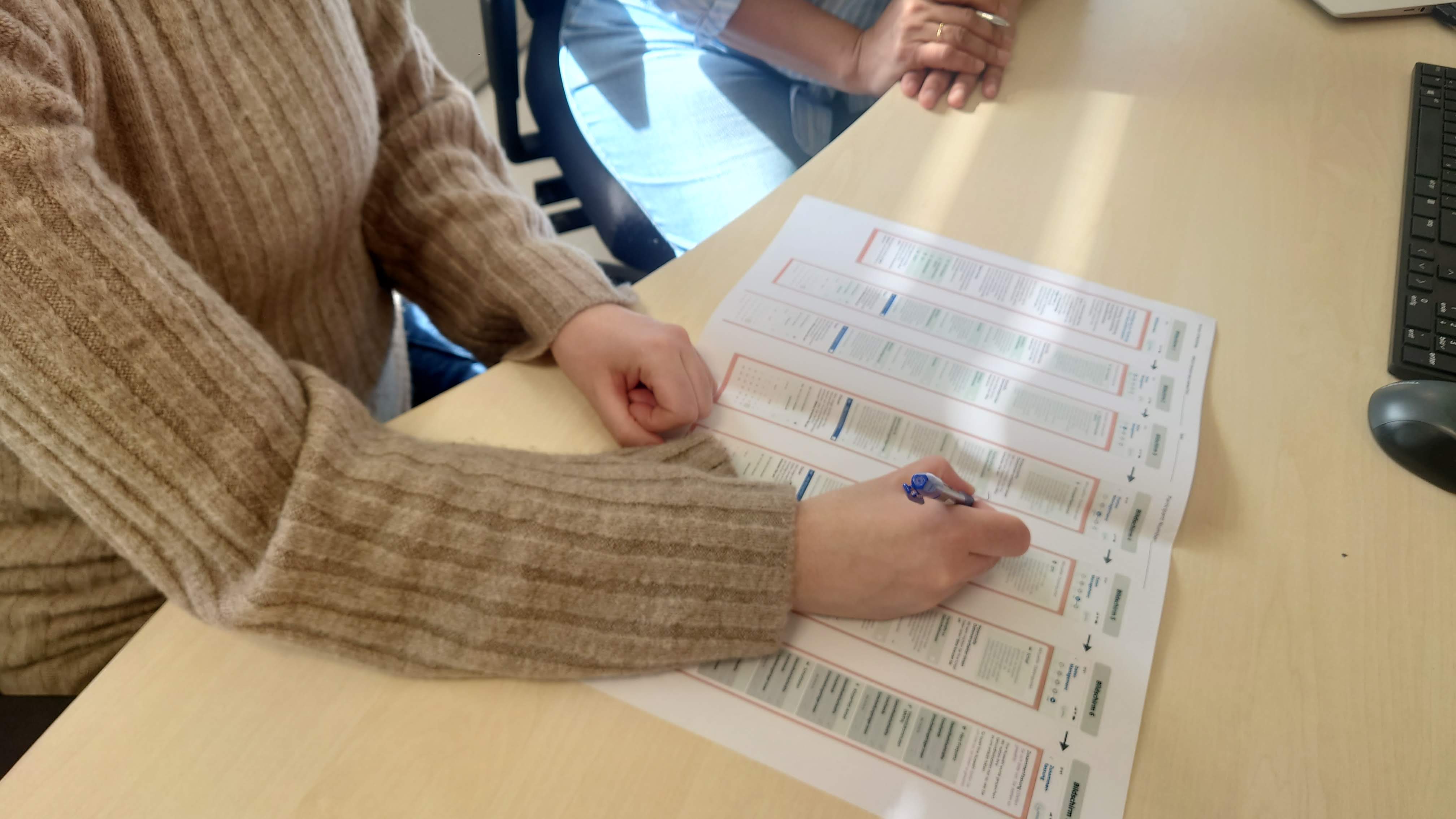}
    \caption{Pilot participant annotating on a printed prototype}
    \Description{An image of a participant annotating on a printed prototype' showing a person sitting at a light tan wooden desk. The person, wearing a ribbed brown sweater, holds a blue pen in their right hand and is annotating a printed sheet of paper laid flat on the desk. The paper contains printed wireframe flows of the multi-step mobile interface screens. In the background across the desk, another seated person's arms and hands are partially visible. A black computer keyboard and mouse sit on the right side of the desk.}
    \label{fig:participantAnnotate}
\end{figure}

\FloatBarrier
\section{Appendix: Detailed Participant Quotes by Theme}
\label{appendix:quotesThemes}

\begin{longtable}{p{0.10\textwidth} p{0.8\textwidth}}
\caption{Thematic analysis and supporting participant quotes.} \label{tab:quotes} \\
\toprule
\textbf{Participant} & \textbf{Quote} \\
\midrule
\endfirsthead

\multicolumn{2}{c}%
{{\bfseries \tablename\ \thetable{} -- continued from previous page}} \\
\toprule
\textbf{Participant} & \textbf{Quote} \\
\midrule
\endhead

\midrule
\multicolumn{2}{r}{{Continued on next page}} \\
\bottomrule
\endfoot

\bottomrule
\endlastfoot

\multicolumn{2}{l}{\textbf{Theme: Control Burden Paradox}} \\
\midrule
P2 & I would choose [Single Step] It’s simpler, clearer. [Multi-step] is more detailed, but more confusing. Since there are more options, [Multi-step] approach could yield better results (B) than the [Single Step], which tend to drag on. It also requires time and attention, especially at our age. \\
\addlinespace
P4 & It seems [the Multi-step version] is more detailed to me somehow. The [Single Step] version is probably faster. Yeah, if I'm taking part in something like that, I certainly don't worry about the exact [time]. \\
\addlinespace
P5 & [Prefer Single step] - Shorter, more concise. \\
\addlinespace
P6 & Yes, I keep thinking about the time. How long will it take me to fill this out? Would I not do this now because it’s too much? Or would I do it now because it’s short and clear? ... If there were even more, I’d start to panic for [the Multi-step version]. That would be too much data for me ... I’d like something in between [the Multi-step version and Single Step] best. I’m also always interested in going a little deeper. \\
\addlinespace
P7 & I liked the [Multi-step version] better... Because if I'm unfamiliar with it, I want to be able to select each individual point once, do I want this, do I not want this ... it always depends, there are people who find learning difficult, and some who need longer to learn... If I have someone who understands that and wants to be done with it, who wants to do it in a flash, then we don't need to discuss it. \\

\midrule
\multicolumn{2}{l}{\textbf{Theme: Contextual Granularity}} \\
\midrule
P1 & Health Care professionals are fine with me, research anyway. And with the decision-makers, it is the question of which ones. Although I say that, of course, especially as a decision-maker, and we hear that if it's anonymized here, it's certainly a positive story. \\
\addlinespace
P3 & government decision-makers ... [should not] receive any personal data, but rather just some anonymized analysis results. \\
\addlinespace
P3 & I think I would actually grant a kind of general authorization for [clinical research studies], more or less. As long as it serves or helps to change or improve other clinical fields, and potentially help those affected, through insights and such studies, I think I’d be on board. \\
\addlinespace
P4 & The question is specifically about insurance companies and government agencies. For me personally, that would be a place where the data doesn’t belong... To me, government agencies simply shouldn’t have access to health data specifically about my health my weight, my blood pressure, and how fit I am. That’s certainly data that’s of interest to insurance companies. For example, whether there’s a difference if I weigh 120 kilos or maybe just 80 kilos. \\
\addlinespace
P4 & When I make my data available to the [research institute], I trust that they will simply handle it with care. \\
\addlinespace
P5 & It wouldn't be a bad idea if there were an option to check a box for each [data] individual category. It’s too general like this. And then I’d have the option to go into a bit more detail. \\
\addlinespace
P7 & [Sharing health data] With a doctor, yes. My cardiologist knows my goals and says, for example, yes, we can do that. No, we can't do that. Then I stick to it. \\
\addlinespace
P7 & So for me, it would be the [Multi-step version]. For me. Because I take enough time for it, because I'm interested in it. If I now say that I'm less interested in it, then of course it's the concise version. \\

\midrule
\multicolumn{2}{l}{\textbf{Theme: Saliency of AI controls}} \\
\midrule
P1 & I assume that AI is already running in the background of these apps. Not really, because I assume that as soon as IT is involved and everything is entered, AI is already working in the background. \\
\addlinespace
P2 & So, actually, wherever I’ve checked the boxes for specialists and researchers, it’s already assumed that you’ll continue with artificial intelligence. \\
\addlinespace
P4 & The problem is that you probably don’t even know where [AI] comes into play anymore. I think that’s the bigger problem... Yes, it’s not something we can stop anyway.. there’s no way around it. \\
\addlinespace
P5 & I’d say to include [AI usage] as a separate category overall. Not just for researchers, but everywhere. \\
\addlinespace
P6 & Well, if it’s noted that this data is processed using artificial intelligence, then I’d put that right at the beginning. \\
\addlinespace
P7 & [AI usage] should be a global point. \\
\end{longtable}

\end{document}